\documentclass[reviewcopy]{elsart}

\usepackage{amsmath}
\usepackage{amsfonts}
\usepackage{amssymb}
\usepackage [ansinew] {inputenc}
\usepackage[dvips]{graphicx}
\usepackage[nomarkers,nolists]{endfloat}
\usepackage[square, comma, sort&compress]{natbib}

\usepackage{theorem}
\theorembodyfont{\upshape}

\newcommand{\ekx} {
\ensuremath{\mathcal{X}}
}

\begin{document}

\begin{frontmatter}
\title{Microcanonical foundation of nonextensivity and generalized thermostatistics
based on the fractality of the phase space}
\author{Vladimir~Garc\'{\i}a-Morales\corauthref{cor1}}
\ead{vladimir.garcia@uv.es},
\author{Julio~Pellicer} \corauth[cor1]{Corresponding author. Tel: +34
96 354 3119; Fax: +34 96 354 3385}
\address{Departament de Termodin\`{a}mica, Universitat de
Val\`{e}ncia, C/Dr. Moliner 50, E-46100 Burjassot, Spain}
\begin{abstract}
\noindent We develop a generalized theory of (meta)equilibrium
statistical mechanics in the thermodynamic limit valid for both
smooth and fractal phase spaces. In the former case, our approach
leads naturally to Boltzmann-Gibbs standard thermostatistics
while, in the latter, Tsallis thermostatistics is
straightforwardly obtained as the most appropriate formalism. We
first focus on the microcanonical ensemble stressing the
importance of the limit $t \to \infty$ on the form of the
microcanonical measure. Interestingly, this approach leads to
interpret the entropic index $q$ as the box-counting dimension of
the (microcanonical) phase space when fractality is considered.
\end{abstract}
\end{frontmatter}

\section{Introduction}
\label{intro}

Recent years are witnessing a growing interest on foundational
aspects of statistical mechanics and thermodynamics in an effort
to extend methods of standard Boltzmann-Gibbs (BG)
thermostatistics well beyond their traditional domain of
application. Systems exhibiting long-range interactions
(self-gravitating systems, two dimensional vortices, metallic
clusters, etc.), fractal memory, spatio-temporal chaos, non
exponential relaxation and non maxwellian velocity distributions,
pose formidable problems to standard BG thermostistics or just
make it unapplicable. Experimental signatures of these systems
include, among others, negative specific heats, inequivalence of
statistical ensembles, phase transitions and self-organization.

Because of the great success of BG thermostatistics in explaining
and predicting thermal properties for a wide class of systems,
this formalism is a cornerstone in which any generalization should
be inspired and to which it has to be reduced under certain
conditions. What are these conditions and the most suitable points
to introduce generalizations in the standard theory have been
issues of intense debate and controversy. Theories like Hill
Nanothermodynamics \cite{HillWorks} proceed from standard
thermodynamics extending the Gibbs-Duhem equation very much in the
manner in which Gibbs extended the former formalism of
thermodynamics giving rise to the extremely important concept of
\emph{chemical potential}. Other generalizations introduce
nontrivial insights on the standard formalism, as the Hessian of
the microcanonical entropy in studying phase transitions in small
systems \cite{Gross} (rejecting statistical ensembles other than
the microcanonical out ot the thermodynamic limit). Another quite
original approach due to Tsallis \cite{Tsallis88,Tsallis98,web}
gave rise in 1988 to an entirely new field in Statistical
Thermodynamics focused on the concept of entropy and its
appropriate form to deal with complex systems. We have shown
recently how Tsallis Thermodynamics can be connected to Hill's
Nanothermodynamics \cite{GarciaMorales05}.

Tsallis thermostatistics has received wide attention recently,
proving quite useful in the analysis of nonlinear low dimensional
maps and multifractals \cite{multifractals} besides other many
applications including: stellar polytropes \cite{Plastino93}, Levy
distributions \cite{Levy}, anomalous diffusion \cite{diffusion},
inelasticity in hadronic reactions \cite{Navarra03}, long-range
interactions \cite{Long}, ionic solutions \cite{GarciaMorales04}
and fully developed turbulence \cite{turbulence}. If we consider a
continuum spectrum of configurations $\omega \in \Omega$
($d\omega= \prod_{i}dp_{i}dq_{i}$) where $\Omega$ is the phase
space volume of the system, Tsallis entropy has the following form
\begin{equation}
S_{q}=k_{q}\frac{\int_{\Omega}\rho^{q}(\omega)d\omega-1}{1-q}
\qquad (q \in \Re) \label{qentropy}
\end{equation}
where $q$ is the so-called entropic parameter. It is easy to see
that when $q$ tends to unity, the well-known Gibbs-Shannon entropy
$S_{1}=-k_{1}\int_{\Gamma}\rho(\omega)\ln \rho(\omega)d\omega$ is
regained.

When equiprobability for the $\Gamma$ microstates of the phase
space is considered, Tsallis entropy takes the form
\begin{equation}
S_{q}^{(mic)}=k_{q}\frac{\Gamma^{1-q}-1}{1-q} \label{TsalMic}
\end{equation}
which is the generalization of the Boltzmann microcanonical
entropy (when $q \to 1$, the Boltzmann entropy
$S_{1}^{(mic)}=k_{1}\ln \Gamma$ is regained).

Maximization of Tsallis entropy, Eq. (\ref{qentropy}), through
appropriate constraints leads to q-exponential distributions.
These are of great interest to study the power laws which many
complex systems exhibit. Tsallis q-exponential distributions have
been extensively founded through rigorous lines \cite{Rajagopal02}
and have the following form
\begin{equation}
p_{i} \propto \left(1-(1-q)\beta E_{i}\right)^{\frac{1}{1-q}}
\end{equation}
When $q$ tends to unity the Boltzmann factor is regained.

The methodology to deal with complex systems is, however, far from
being a closed debate. Other interesting generalizations have
appeared recently based in the use of deformed logarithms or kappa
distributions \cite{Kaniadakis1, Kaniadakis2}, Kolmogorov-Nagumo
averaging \cite{NaudtsWorks} and Renyi statistics
\cite{Baskhirov}. A plethora of new entropies have also appeared
during the last decade, many of them inspired in the Tsallis
MaxEnt approach \cite{NoteKaniadakis}.

Although Tsallis entropy has many desirable mathematical
properties such as concavity and Lesche stability \cite{AbeWorks},
besides a uniqueness theorem \cite{Santos} having also associated
an H-theorem \cite{Kaniadakis1, PlastinoPRL}, the rationale behind
its form and why it is so suitable to deal with the thermodynamics
of complex systems against other entropic forms has not been
clarified. Tsallis introduced his entropy inspired in the study of
multifractals. The entropy appears in a postulate-like manner in
his first work on the subject \cite{Tsallis88} and, although its
interest and beautiful properties (pseudoadditivity, concavity for
$q>0$, etc.) are soon apparent, it would be desirable to have an
approach giving some insight in the starting point, i.e. the form
of the entropy, which is the building block of the formalism.

In this paper such an approach is attempted using concepts in
which Tsallis himself has been insisting since his above mentioned
proposal. The crucial issues are: the infinite time limit,
nonergodicity and fractality of the phase space. When the latter
is thermodynamically important, the phase space is not a smooth
surface and standard thermostatistics does not apply
\cite{Gallavotti99}. Although the existence of fractality in the
phase space of many-body complex systems far from equilibrium is
an issue under debate there is plenty of evidence in low
dimensional systems (see for example \cite{Hoover98} where the
nice example of the Galton board is discussed). We show here that
fractal analysis of the phase space leads quite naturally to
Tsallis entropy. The approach allows us to give an interpretation
of the entropic parameter $q$ in terms of the phase space geometry
adding insight to previous findings
\cite{GarciaMorales05,turbulence,qmeaning} in which $q$ was
related to thermal variables and/or their fluctuations.

Although we are concerned also with ensemble theory, our reasoning
line here is strictly based on the microcanonical ensemble and on
how the findings in this ensemble translate to the other ones by
means of the Legendre transform mechanism. In this study
Massieu-Planck entropic potentials play an important role in
generalizing our results to other ensembles. The interest of these
quantities at nonequilibrium problems were recognized at the very
moment of their inception more than one century ago
\cite{Massieu}. These allow for a compact formulation of
generalized statistical mechanics and thermodynamics (see
\cite{Vives} for a Jaynes-like formulation of equilibrium standard
thermodynamics by using these entropic potentials).

\section{Microcanonical measures and ergodicity}

Let us consider a system of $N$ particles isolated from its
surroundings by rigid, impermeable, adiabatic walls. The
trajectory of the motion is made in such a way that energy is
constant at all times (provided that the system is perfectly
isolated). Let us consider a phase space with volume $\Omega$
containing all possible configurations that the system can attain.
Each of these configurations is a cell of volume $\sim h^{3N}$ in
which the phase space is subdivided. If we want to count the total
number of cells visited by the \emph{actual} trajectory of the
motion during a time $T$ this can be achieved \emph{exactly} by
means of the following \emph{mechanical} quantity
\begin{eqnarray}
&&\int_{0}^{T}dt\int_{\Omega}\prod_{i}\frac{dp_{i}dq_{i}}{h^{3N}}\delta\left(H(\omega)-E\right)
|\dot{Q}_{i}||\dot{P}_{i}|\delta\left(p_{i}-P_{i}\right)\delta\left(q_{i}-Q_{i}\right)
\label{jumps}
\end{eqnarray}
where $Q_{i}(t)$ and $P_{i}(t)$ are, respectively, the generalized
position and momenta coordinates which are the solution of the
$6N$ Hamilton equations of motion
\begin{eqnarray}
\dot{Q}_{i}(t)=\frac{\partial H}{\partial P_{i}} \nonumber \\
\dot{P}_{i}(t)=-\frac{\partial H}{\partial Q_{i}} \label{micro}
\end{eqnarray}
The phase space visited by the trajectory is thus \emph{embedded}
in the available phase space $\Gamma$ (the constant energy surface
in the volume $\Omega$). Note that in Eq.(\ref{jumps}), the first
Dirac delta containing the difference $H(p,q)-E$ is redundant if
in the actual trajectory specified by $(P, Q)$ energy is
conserved. It is written here, however, for our convenience,
emphasizing that the trajectory lies in the constant energy
surface. Not all points in the constant energy surface, however,
are actually points of the trajectory and this is what the other
delta functions select: only points on the mechanical trajectory
do contribute then to Eq.(\ref{jumps}).

We now define what we are going to call from here on \emph{actual
microstates} $\widetilde{\Gamma}$
\begin{eqnarray}
\widetilde{\Gamma}&=&\lim_{T \to \infty}
\int_{0}^{T}dt\int_{\Omega}\prod_{i}\frac{dp_{i}dq_{i}}{h^{3N}}\delta\left(H(\omega)-E\right)|\dot{Q}_{i}||\dot{P}_{i}|\delta\left(p_{i}-P_{i}\right)\delta\left(q_{i}-Q_{i}\right)
\label{actual}
\end{eqnarray}
We introduce now also the smooth microcanonical measure $\Gamma$
or, what we will call from here on \emph{smooth} or
\emph{Boltzmann microstates}
\begin{equation}
\Gamma=\int_{\Omega}\prod_{i}\frac{dp_{i}dq_{i}}{h^{3N}}\delta\left(H(\omega)-E\right)
\label{smooth}
\end{equation}
We should say that the system is \emph{ergodic} if
$\widetilde{\Gamma}=\Gamma$. It is clear that the system will be
ergodic if
\begin{equation}
\lim_{T \to
\infty}\int_{0}^{T}dt\prod_{i}|\dot{Q}_{i}||\dot{P}_{i}|\delta\left(p_{i}-P_{i}\right)\delta\left(q_{i}-Q_{i}\right)=1
\label{cond1}
\end{equation}
for each point $\omega=(\mathbf{q},\mathbf{p}) \in \Gamma$. This
means that each point of the attainable phase space (with
$H(\mathbf{q},\mathbf{p})=E=constant$) becomes, in fact, a locus
of the trajectory in the long time limit, i.e. the trajectory
fills densely the constant energy surface. It is important to note
that writing here the limit $T$ tending to infinity contains a bit
of abuse of language since time must be, in any case, lower than
the Poincaré recurrence time to make sure that each phase space
cell is not visited more than once during the trajectory. The
limit $T \to \infty$ should be understood as a long time enough to
make of a moving point in the phase space (a \emph{microscopic}
description of the system) a trajectory which is long enough to be
considered itself as a \emph{macroscopic} geometric object (if we
think in a many particle system at finite temperatures). This is
the key idea to understand our approach and is applied when the
trajectory has achieved its maximum spread over the phase space
surface. We then do not care about of in which phase space point
of the trajectory is actually the system: the macroscopic
trajectory becomes the key geometrical object to pass to a
statistical description from the microscopic mechanics. We then
attribute the physical properties of the motion of the system
(\emph{mechanically}, a moving point in a high number of
dimensions) to an average property of the whole trajectory
(\emph{statistically}, a complex geometrical structure embedded in
the surface of constant energy).

Let us see that in the case of an ergodic system the time average
(mechanical average) and the ensemble average (i.e. over the
smooth microcanonical measure) of a phase function
$f(\mathbf{P}(t),\mathbf{Q}(t))$ coincide. We define the long time
average of $f$, over the trajectory $<f>$ as
\begin{eqnarray}
&&<f>=\frac{1}{\widetilde{\Gamma}}\lim_{T \to
\infty}\int_{0}^{T}dt
\int_{\Omega}\prod_{i}\frac{dp_{i}dq_{i}}{h^{3N}}\delta\left(H(\omega)-E\right)\times
\nonumber \\
&& \times
f(\mathbf{P}(t),\mathbf{Q}(t))|\dot{Q}_{i}||\dot{P}_{i}|\delta\left(p_{i}-P_{i}\right)\delta\left(q_{i}-Q_{i}\right)
\label{timeav}
\end{eqnarray}
Because of the product of Dirac deltas in the time dependent
integrand, we have $f(\mathbf{P}(t),\mathbf{Q}(t))=f(\omega)$, so
that $f$ can go out of the time integral. Applying
Eq.(\ref{cond1}) and $\widetilde{\Gamma}=\Gamma$ we obtain
\begin{equation}
<f>=\frac{1}{\Gamma}\int_{\Omega}\prod_{i}\frac{dp_{i}dq_{i}}{h^{3N}}f(\omega)\delta\left(H(\omega)-E\right)=\bar{f}
\label{Birk}
\end{equation}
i.e. the time average of the phase function $f$ equals the
ensemble average $\bar{f}$.

Ergodicity is the mechanism that allows us to pass from a
\emph{mechanical} description of the system to a
\emph{statistical} one. We can wonder what should we do if the
system is not ergodic i.e. if $\widetilde{\Gamma} \ne \Gamma$. It
is clear, in any case, that the measure $\widetilde{\Gamma}$ is
embedded in the smooth measure $\Gamma$ composed of all
microstates in the constant energy surface. If we can still find
$\widetilde{\Gamma}$ as a function of the embedding smooth surface
$\Gamma$ in a form $\widetilde{\Gamma}=\mathcal{N}(\Gamma)$, we
can still attempt a relatively simple statistical description. In
what follows, we will focus in this smooth surface considering
physical cases in which the function $\mathcal{N}(\Gamma)$ can be
constructed. For the ergodic case, $\mathcal{N}(\Gamma)=\Gamma$.
If the system is not ergodic, we can still have a relatively
simple statistical description of it if the trajectory is a
fractal curve embedded in the smooth energy surface of $\Gamma$
cells. If the reduced box-counting dimension of such a curve is
$d$ $( \equiv D/6N$ so that $ 0 \le d \le 1)$ and provided that
the volume $\Delta$ ($\propto 1/\Gamma$) of each phase space cell
is vanishingly small, we have \cite{Falconer85, Beck93}
\begin{equation}
\mathcal{N}(\Gamma) \propto \Gamma^{d}
\end{equation}
The available phase space is now a fractal object embedded in the
Boltzmann $6N$-dimensional smooth phase space. Clearly, when $d=1$
(6N-dimensional phase space) we regain ergodicity.

\section{General definition of the microcanonical entropy
and particularization for both smooth and fractal phase spaces}

The first step in passing from mechanics to statistics on the
phase space has been completed in the previous section. The
relationship to thermodynamics has to emerge now to have thermal
variables founded microscopically. In doing that we can make the
following (necessarily heuristic) reasoning. In a nonequilibrium
situation, the \emph{change} in the number of smooth microstates
(i.e. on the area of the smooth energy surface) can be thought to
be directly proportional to the number of \emph{actual}
microstates of the trajectory. This can be clearly understood if
we think on each point of the actual trajectory as a \emph{source}
for new smooth microstates. It is in this way that our \emph{lack
of knowledge} on the microstate in which the system is can spread,
tending to increase. \emph{The smooth energy surface and the
actual trajectory of the system which spreads on it are
inextricably related to each other.} Since we do not know indeed
in which microstate is the system, we are forced to admit that
each microstate behaves in nonequilibrium as a source for new ones
and these new microstates make also the smooth surface to
increase. All these considerations can be written mathematically
as follows
\begin{equation}
\frac{d \Gamma}{dt}=\sigma \mathcal{N}(\Gamma) \label{GammaT}
\end{equation}
Here $\sigma$ is the rate of smooth microstates production from
the actual $mechanical$ ones. It is exactly at this point that we
introduce thermodynamics by establishing the equivalence between
this rate and that of entropy production.
\begin{equation}
\sigma=\frac{d \mathcal{S}}{dt} \label{sigma}
\end{equation}
Here $\mathcal{S}=S/k$ is the dimensionless entropy. If we replace
Eq.(\ref{sigma}) in Eq. (\ref{GammaT}) we obtain
\begin{equation}
\frac{d \mathcal{S}}{d \Gamma}=\frac{1}{\mathcal{N}(\Gamma)}
\qquad \Rightarrow \qquad
\mathcal{S}=\int_{1}^{\Gamma}\frac{d\Gamma'}{\mathcal{N}(\Gamma')}
 \label{entropy}
\end{equation}
where we have considered in the integration that for
$\Gamma_{0}=1$ the entropy vanishes. This constitutes a general
definition of entropy that, as we are going to see, reduces to
well known entropic forms after considering models for
$\mathcal{N}(\Gamma)$. Let us consider first the ergodic case
(smooth phase space with $\mathcal{N}(\Gamma)=\Gamma$ as discussed
in the previous section). From our definition we obtain the
Boltzmann entropy after integration
\begin{equation}
\mathcal{S}=\ln \Gamma
\end{equation}
If, however, a fractal phase space is considered
($\mathcal{N}(\Gamma) \propto \Gamma^{d}$), we have, after
integration (absorbing the proportionality constant in a
generalized Boltzmann constant $k_{d}$ so that
$\mathcal{S}=S/k_{d}$)
\begin{equation}
\mathcal{S}=\frac{\Gamma^{1-d}-1}{1-d}
\end{equation}
Quite remarkably, this form of the microcanonical entropy is
similar to that of Tsallis, Eq. (\ref{TsalMic}) under the
correspondence $q \equiv d$. If this connection is made, our view
implies that \emph{the entropic parameter $q$ is equal to the
reduced box-counting dimension $d=D/6N$ of the available fractal
phase space}. Of course, for Boltzmann systems, $D=6N$, $d=1$,
$q=1$ and the Boltzmann entropy is regained. These considerations
are relevant to many physical systems due to the multifractality
that thermostatted and low dimensional systems exhibit
\cite{Hoover98} (here we are considering isolated systems in the
microcanonical ensemble and, therefore, we are focusing in the
fractal support of the phase space).

This insight allows us to interpret unequivocally the strong
coupling regime in Ref. \cite{GarciaMorales04} found in ionic
solutions. In \cite{GarciaMorales04} we observed that $q$ tends to
vanish in this limit. It was previously known that when
multivalent ions of the same charge are located close to a highly
charged surface these crystallize forming a Wigner crystal
\cite{Netz01}. This means that the available classical phase space
collapses into regions with strikingly lower dimensions.
Therefore, $D<<6N$, $d \to 0$ and $q \to 0$, as previously
obtained in \cite{GarciaMorales04} because of the crystalline
ordering at the interface. In general, for moderate coupling,
there is quasicristallyne behavior at the interface which is still
markedly different to that in the bulk (a disordered, uncorrelated
phase). The transition between the phase space for the bulk and
the interface can, hence, be depicted as a complex filamentary
structure having fractal properties. At low coupling values, the
liquid in the interface has bulk properties \cite{Netz01}, and
both, bulk and interfacial liquid merge into a single homogeneous
and smooth phase. All this variety of behavior is explained
economically through the index $q$. The weak coupling limit can
also be understood analytically following a previous insight in
which we related $q$ to the interfacial and bulk entropies
\cite{GarciaMorales05}.

\section{Generalized canonical ensemble}

The above methodology can be easily extended to the canonical
ensemble provided that we can characterize the space in which the
typical trajectory is embedded by a number of configurations
$\mathcal{N}(\Gamma^{*})$ where $\Gamma^{*}$ now denotes the total
number of smooth configurations in the composite phase space
surface appropiately weighted by the different energies that the
different regions of the surface have. Following the same
principle for the evolution of the number of microstates (now
weighted) we have
\begin{equation} \frac{d \Gamma^{*}}{dt}=\sigma^{*} \mathcal{N}(\Gamma^{*})
\label{GammaT2}
\end{equation}
where
\begin{equation}
\sigma^{*}=\frac{d(-F/kT)}{dt} \label{sigma2}
\end{equation}
is the rate of change of the natural entropic potential which in
the canonical ensemble is minus the Helmholtz free energy $F$ in
$kT$ units. This entropic potential is related to entropy by means
of the Legendre transform mechanism
\begin{equation}
-\frac{F}{kT}=\mathcal{S}-\frac{E}{kT} \label{Legendre}
\end{equation}
Note that the natural entropic potential correctly accounts for
the entropic contribution of the phase space cells substracting to
them the energy contribution which is now a fluctuating variable.
If we define $\mathcal{F}\equiv -F/kT$ and replace
Eq.(\ref{sigma2}) in Eq.(\ref{GammaT2}) we obtain
\begin{equation}
\frac{d \mathcal{F}}{d
\Gamma^{*}}=\frac{1}{\mathcal{N}(\Gamma^{*})} \qquad \Rightarrow
\qquad \mathcal{F}
=\int_{1}^{\Gamma^{*}_{f}}\frac{d\Gamma^{*}}{\mathcal{N}(\Gamma^{*})}
 \label{entropyP}
\end{equation}
where we have used that for $\Gamma^{*}_{0}=1$ the entropic
potential vanishes. For an ergodic system, despite that now
regions of the phase space are weighted by their differing energy,
all them can be attained in the long-time limit so that
$\mathcal{N}(\Gamma^{*})=\Gamma^{*}$ and
\begin{equation}
\mathcal{F}=\ln \Gamma^{*}
\end{equation}
The total number of ``weighted microstates'' $\Gamma^{*}$ is the
usually called \emph{partition function} of the system. For a
fractal phase space, although the microstates are weighted by
energy, the microcanonical restriction coming from a lower
dimension of the actual phase space still applies and we have
after integration
\begin{equation}
\mathcal{F}=\frac{\Gamma^{*1-d}-1}{1-d}
\end{equation}
Our aim now is to calculate the weights in the microstates
$\Gamma^{*}$ to relate them to regions in which the phase space is
partitioned. It is important that if subindex $i$ denotes regions
of constant energy $E_{i}$ we have
\begin{eqnarray}
\Gamma^{*}&=&\sum_{i}\Gamma^{*}_{i} \label{partition}\\
\Gamma&=&\sum_{i}\Gamma_{i}
\end{eqnarray}
and if for each of these regions Eq.(\ref{Legendre}) applies
\begin{equation}
\mathcal{F}_{i}=\mathcal{S}_{i}-\beta E_{i} \label{Legendrei}
\end{equation}
Now, since the definitions Eqs. (\ref{entropy})
and({\ref{entropyP}) apply also to $\mathcal{S}_{i}$ and
$\mathcal{F}_{i}$ by replacing these in Eq.({\ref{Legendrei}) we
can calculate $\Gamma^{*}_{i}$ as a funtion of each $\Gamma_{i}$
and the energy $E_{i}$
\begin{equation}
\int_{1}^{\Gamma^{*}_{i}}\frac{d
\Gamma^{*}_{i}}{\mathcal{N}(\Gamma^{*}_{i})}=\int_{1}^{\Gamma_{i}}\frac{d\Gamma_{i}}{\mathcal{N}(\Gamma_{i})}-\beta
E_{i} \label{Gi}
\end{equation}
For example, for an ergodic system we have, from the latter
equation
\begin{equation}
\Gamma^{*}_{i}=\Gamma_{i}e^{-\beta E_{i}}
\end{equation}
and, for the nonergodic system with fractal phase space
($\mathcal{N}(\Gamma^{*}_{i})=\Gamma^{*d}_{i}$) we have
\begin{equation}
\Gamma^{*}_{i}=\Gamma_{i}\left(1-(1-d)\beta
E_{i}/\Gamma_{i}^{1-d}\right)^{\frac{1}{1-d}}
\end{equation}
which coincides with the Tsallis $q-exponential$ form with the
only difference that $\beta E_{i}$ is here divided by the
degeneration of the microstate. Note, however, that usually the
MaxEnt approach requires the specification of the entropic form as
well as the appropriate constraints. Here we have not chosen, in
principle, any entropic form, nor have privileged any choice for
the constraints: we have based all our reasoning in the fractal
analysis of the phase space and in the evolution equation for the
number of ''smooth'' microstates.

The partition functions for each system can now be calculated from
Eq.(\ref{partition}). The probability of having the system with
energy $E_{i}$ is given in each case by the ratio
$\Gamma^{*}_{i}/\Gamma^{*}$.

\section{Massieu-Planck thermodynamic potentials and generalized thermostatistics}

In the previous sections we have formulated equilibrium
generalized thermostatistics in the canonical and microcanonical
ensembles. In general, the motion is such that a set of extensive
variables $X_{j}$ are kept constant during the trajectory (while
fluctuating their conjugate intensive ones $y_{j}$) and the
intensive variables $y_{k}$ are also kept constant (and then their
extensive conjugate ones $X_{k}$ fluctuate). This always occurs in
semi-open systems in which some extensive variables fluctuate. We
can now define a generalized thermodynamic potential $\ekx$ by
Legendre transforming the entropy, following a totally analogous
procedure to the sketched above for the canonical ensemble
\begin{equation}
\ekx=\mathcal{S}-\sum_{k}y_{k}X_{k} \label{defekx}
\end{equation}
This is the thermodynamical definition of the generalized
potential $\ekx$. Its character of thermodynamic representation is
made explicit in its generalized differential Gibbs form
\begin{equation}
d\ekx=\sum_{j}y_{j}dX_{j}-\sum_{k}X_{k}dy_{k}
\end{equation}
from which it is seen that non-environment variables $y_{j}$ and
$X_{k}$ can be obtained from the natural ones by differentiation
\begin{equation}
X_{k}=\frac{\partial \ekx}{\partial y_{k}} \qquad
y_{j}=-\frac{\partial \ekx}{\partial X_{j}} \label{macro}
\end{equation}
by keeping constant all other variables not involved in the
differentiation. In Table \ref{Table1}, the sets of variables and
the form of $\ekx$ for each statistical ensemble are indicated.

The typical long-time trajectory contains now
$\mathcal{N}(\Upsilon)$ microstates where the smooth microstates
in the constant $X_{j}$'s surface $\Upsilon$ obbey the following
evolution equation when equilibrium is perturbed
\begin{equation}
\frac{d \Upsilon}{dt}=\dot{\ekx}\mathcal{N}(\Upsilon) \label{Upsi}
\end{equation}
which leads to a microscopic definition of $\ekx$ (in a way
totally analogous to previous sections) as
\begin{equation}
\ekx=\int_{1}^{\Upsilon}\frac{d \Upsilon'}{\mathcal{N}(\Upsilon')}
 \label{ekxst}
\end{equation}
and we also have from Eq.(\ref{defekx}) particularized for a
micro-value $i$ of each extensive variable $X_{k}$ (denoted by
$X_{k,i}$)
\begin{equation}
\int_{1}^{\Upsilon_{i}}\frac{d
\Upsilon_{i}'}{\mathcal{N}(\Upsilon_{i}')}=\int_{1}^{\Gamma_{i}}\frac{d\Gamma_{i}}{\mathcal{N}(\Gamma_{i})}-\sum_{k}y_{k}X_{k,i}
\label{GiGen}
\end{equation}
which allows to know $\Upsilon_{i}$ as a function of all the
weights and then evaluating the generalized partition function
from $\Upsilon=\sum_{i}\Upsilon_{i}$ (where is to be understood
that the number of sums is the same than that of fluctuating
extensive variables). For an ergodic system, for which
$\mathcal{N}(\Upsilon_{i})=\Upsilon_{i}$, this leads to the
generalized Gibbs distribution
\begin{equation}
\Upsilon=\sum_{i}\Gamma_{i}e^{-\sum_{k}y_{k}X_{k,i}}
\end{equation}
and, in the non-ergodic case with fractal phase space
($\mathcal{N}(\Upsilon_{i})=\Upsilon_{i}^{d})$
\begin{equation}
\Upsilon=\sum_{i}\Gamma_{i}\left(1-(1-d)\sum_{k}
y_{k}X_{k,i}/\Gamma_{i}^{1-d}\right)^{\frac{1}{1-d}}
\end{equation}

\section{Conclusions}

The unifying principle of the smooth phase space evolution Eq.
(\ref{Upsi}) allows to build a generalized thermostatistic
formalism concerning macrovariables analogously to that of
Hamiltonian mechanics containing microvariables (comparison of
Eqs. (\ref{micro}) with (\ref{macro}) makes explicit this analogy:
the role played by $\ekx$ in macroscopic systems is quite similar
to the one played by $H$ in microscopic ones). It is to be noted
that this principle embodies the maximum entropy principle in the
microcanonical ensemble as well as the minimum of the conventional
thermodynamic potentials at equilibrium (a maximum of their
closely related Massieu-Planck entropic potentials, see Table
\ref{Table1}) for each ensemble. This is clearly seen in the
differential equation which serves as definition of the entropy
Eq. (\ref{entropy}), since in the microcanonical ensemble
$1/\mathcal{N}(\Gamma)$ is directly related to probability. As
probability is a positive quantity which decreases with increasing
the available phase space (of course, this applies far from phase
separation) entropy is here, therefore, a \emph{concave and
increasing} function of its argument. The generalized
Massieu-Planck entropic potential $\ekx$ describes all the
thermodynamics of a given system when known as a function of the
natural variables and is the key stone, in our view, in connecting
the macro to the microworld.

Our principle for the evolution of the smooth surface of
microstates vaguely reminds the ''differential equations''
formulation of q-exponential distributions \cite{Tsallis04}. Here,
however, the formulation does not come from mathematical
motivation and/or developing of mathematical identities, but an
attempt has been made to substantiate it physically.

We have shown that our picture leads naturally to BG statistical
mechanics as well as to Tsallis formalism, without need of
formulating the latter by introducing \emph{ad hoc} constraints.
Tsallis entropy gains advantage in our formulation compared to
other entropic measures since its physical meaning is here
appealing and has far reaching consequences beyond any \emph{ad
hoc} formal construction with whatever nice given properties.

We want to acknowledge B. A. Nikolov and J. A. Manzanares for
their comments and critical reading of a previous version of this
manuscript. V. G-M. thanks also J. M. García Sanchis for
conversations. Financial support from the MCYT (Ministry of
Science and Technology of Spain) and FEDER under project No.
MAT2002-00646 is also gratefully acknowledged.

\begin{table}
\begin{center}
\begin{tabular}{c c c c c c}
\hline
Ensemble & $ \{X_{j}\} $ & $\{y_{k}\}$ & $ \{X_{k}\} $ & $\{y_{j}\}$ & $\ekx$\\
\hline
Microcanonical  & $\{E, V, N\}$ & $\{\emptyset \}$& $ \{\emptyset \} $ & $\{\beta, \pi, \nu\}$ & $\mathcal{S}$\\
Canonical & $\{V, N\}$ & $\{\beta\}$& $ \{E\} $ & $\{\pi, \nu \}$ & $-\beta F$ \\
Grand Canonical & $\{V\}$ & $\{\beta, \nu \}$& $ \{E, N\} $ & $\{\pi\}$ & $-\beta \Xi$ \\
Isothermal-Isobaric  & $\{N\}$ & $\{\beta, \pi \}$& $ \{E, V\} $ & $\{\nu\}$ & $-\beta G$ \\
Open  & $\{\emptyset \}$ & $\{\beta, \pi, \nu \}$& $ \{E, V, N\} $ & $\{\emptyset \}$ & $-\beta \mathcal{E}$\\
\hline
\end{tabular}
\caption{Environment variables $ \{X_{j}\} $ (extensive) and
$\{y_{i}\}$ (intensive) and non-environment fluctuating ones $
\{X_{k}\} $ (extensive) $\{y_{k}\}$ (intensive) for each ensemble
indicated. ($\beta \equiv 1/kT$, $\pi \equiv p/kT$ and $\nu \equiv
-\mu/kT$).} \label{Table1}
\end{center}
\end{table}


\begin{thebibliography}{}

\bibitem{HillWorks}
T. L. Hill, Thermodynamics of Small Systems, Dover, New York,
1994; T. L. Hill, Nano Lett. 1 (2001) 273.
\bibitem{Gross}
D. H. E. Gross, Phys. Chem. Chem. Phys. 4 (2002) 863.
\bibitem{Tsallis88}
C. Tsallis, J. Stat. Phys. 52 (1988) 479.
\bibitem{Tsallis98} C. Tsallis, R. S. Mendes and A. Plastino,
Physica A, 261 (1998) 534.
\bibitem{web}
See http://tsallis.cat.cbpf.br/biblio.htm for an updated
bibliography on the subject.
\bibitem{GarciaMorales05}
V. Garcia-Morales, J. Cervera, J. Pellicer, Phys. Lett. A 336
(2005) 82.
\bibitem{multifractals} M. L. Lyra, C. Tsallis, Phys. Rev. Lett., 80 (1998) 53;
E. P. Borges, C. Tsallis, G. F. J. A\~na\~nos, P. M. C de
Oliveira, Phys. Rev. Lett. 89 (2002) 254103; G. F. J. A\~na\~nos,
C. Tsallis, Phys. Rev. Lett., 93 (2004) 020601.
\bibitem{Plastino93} A. R. Plastino and A. Plastino, Phys. Lett. A, 174 (1993) 384.
\bibitem{Levy} C. Tsallis, S. V. F. Levy, A. M. C. Souza, R. Maynard,
Phys. Rev. Lett., 75 (1995) 3589; D. Prato and C. Tsallis, Phys.
Rev. E, 60 (1999) 2398.
\bibitem{diffusion} D. H. Zanette and P. A. Alemany, Phys. Rev. Lett. 75 (1995) 366;
M. O. Caceres and C. E. Budde, \emph{ibid}. 77 (1996) 2589; D. H.
Zanette and P. A. Alemany, \emph{ibid}. 77 (1996) 2590.
\bibitem{Navarra03} F. S. Navarra, O. V. Utyuzh, G. Wilk, Z. Wlodarczyk, Phys. Rev. D, 67 (2003) 114002.
%\bibitem{Torres97}   D. F.  Torres, H. Vucetich, A. Plastino, Phys. Rev. Lett. 79 (1997) 1588.
\bibitem{Long} R.  Salazar, R. Toral, Phys. Rev.  Lett.,
83 (1999) 4233; S. A. Cannas, F. A. Tamarit, Phys. Rev. B, 54
(1996) R12661.
\bibitem{GarciaMorales04}
V. Garcia-Morales, J. Cervera, J. Pellicer, Physica A 339 (2004)
482.
\bibitem{turbulence} C. Beck, Phys. Rev. Lett., 87 (2001) 180601.
\bibitem{Rajagopal02} A. K. Rajagopal and S. Abe, \emph{Statistical mechanical foundations
for systems with nonexponential distributions} in \emph{Classical
and Quantum complexity and Nonextensive Thermodynamics}, eds. P.
Grigolini, C. Tsallis and B. J. West, Chaos, Solitons and
Fractals, vol. 13, p. 529, Pergamon-Elsevier, Amsterdam, 2002.
\bibitem{Kaniadakis1}
G. Kaniadakis, Physica A 296 (2001) 405, cond-mat/0103467.
\bibitem{Kaniadakis2}
G. Kaniadakis, Phys. Rev. E 66 (2002) 056125.
\bibitem{NaudtsWorks}
J. Naudts, Physica A 340 (2004) 32; J. Naudts, Physica A 332
(2004) 279.
\bibitem{Baskhirov}
A. G. Bashkirov, Phys. Rev. Lett. 93 (2004) 130601.
\bibitem{NoteKaniadakis}
A useful compilation of references on new entropies is given in
the introduction of Ref. \cite{Kaniadakis1}
\bibitem{AbeWorks}
S. Abe, Phys. Rev. E 66 (2002) 046134; S. Abe, Continuum Mech.
Thermodyn. 16 (2004) 237.
\bibitem{Santos}
R. J. V. dos Santos, J. Math. Phys. 38 (1997) 4104.
\bibitem{PlastinoPRL}
J. A. S. Lima, R. Silva, A. R. Plastino, Phys. Rev. Lett. 86
(2001) 2938.
\bibitem{Gallavotti99} G. Gallavotti, Statistical Mechanics. A short treatise
Springer Verlag, Berlin, 1999.
\bibitem{Hoover98}
W. G. Hoover, J. Chem. Phys. 109 (1998) 4164.
\bibitem{qmeaning}
G. Wilk and Z. Wlodarczyk, Phys. Rev. Lett. 84 (2000) 2770 (2000);
D. Jiulin, Europhys. Lett. 67 (2004) 893.
\bibitem{Massieu}
M. F. Massieu, Comptes Rendus Acad. Sci. 69 (1869) 858.
\bibitem{Vives}
E. Vives and A. Planes, J. Stat. Phys. 106 (2002) 827.
\bibitem{Falconer85} K. G. Falconer, The Geometry of Fractal Sets
Cambridge University Press, Cambridge, 1985.
\bibitem{Beck93} C. Beck and F. Schl\"ogl, Thermodynamics of chaotic systems
Cambridge University Press, Cambridge, 1993.
\bibitem{Netz01}
R. R. Netz, Eur. Phys. J. E 5 (2001) 557.
\bibitem{Tsallis04} C. Tsallis, Physica D 193 (2004) 3.
\end{thebibliography}
\end{document}